# Finite Volume Model to Study Calcium Diffusion in Neuron Involving $J_{RYR}$, $J_{SERCA}$ and $J_{LEAK}$

Amrita Tripathi and Neeru Adlakha

**Abstract**—Calcium dynamics is the highly responsible for intracellular electrical (action potential) and chemical (neurotransmitter) signaling in neuron cell. The Mathematical modeling of calcium dynamics in neurons lead to the reaction diffusion equation which involves the parameters like diffusion coefficient, free calcium, bound calcium, buffers and bound buffer. Here the parameters like receptors, serca and leak are also incorporated in the model. Appropriate boundary conditions have been framed based on biophysical conditions of the problem. The finite volume method has been employed to obtain the solution. The computer program has been developed using MATLAB 7.11 for the entire problem to compute $Ca^{2+}$ profiles and study the relationships among various parameters.

—————————— ◆ ——————————

## 1 INTRODUCTION

Intracellular calcium is an important second messenger in living cells. The calcium is required for almost every process in human organs like heartbeat, muscles contractions, bones activity and brain functionality etc. $Ca^{2+}$ dynamics [19] is the exchange of $Ca^{2+}$ ions between intracellular $Ca^{2+}$ stores and the cytosol, entering and leaving ions between the cells and binding activity of calcium and calcium binding proteins. The most important calcium binding proteins are itself buffers that are located in $Ca^{2+}$ stores. The binding of calcium molecules with buffer depends on calcium concentration in the cell [7, 13].

The concentration dependent binding of $Ca^2$ to buffers serves as an indicator of the concentration of free calcium concentration in intracellular measurements. The active elements of the exchange processes are channels, receptors like ryanodine receptor (RyR), IP3 receptor etc. serca and leaks in the membranes. Ryanodine receptors mediate the release of calcium ions from the secroplasmic reticulum. They bound the intracellular $Ca^{2+}$ that is stored in the endoplasmic reticulum and other storage devices. SERCA Pumps transport the $Ca^{2+}$ against its electro chemical gradient. Leak receives the $Ca^{2+}$ that comes from pump and is stored in the endoplasmic reticulum (ER). Within the ER $Ca^{2+}$ maintains the high capacity and low efficiency of $Ca^{2+}$ binding proteins. They maintain the balances of $Ca^{2+}$ ions through active (in) and passive (out) process [4].

In the present study an attempt has been made to model the calcium diffusion in neuron cell involving $j_{RYR}$, $j_{LEAK}$ and $j_{SERCA}$. The model has been developed for one dimensional steady state case. The finite volume method [8, 16, 20] has been employed to obtain the solution. A computer program has been developed in MATLAB 7.11 for the problem and simulated on Core i3 processor with 2.13 GHz processing speed, 64-bit machine with 320 GB memory. Numerical values of physiological parameters have been used to study the calcium concentration.

## 2 MATHEMATICAL FOMULATION

The calcium kinetics in neuron is governed by a set of following equation given by [7, 15]:

$$[Ca^{2+}] + [B_j] \underset{k^-}{\overset{k^+}{\rightleftharpoons}} [CaB_j] \quad (1)$$

Where $[B_j]$ and $[CaB_j]$ are free and bound buffers respectively, and 'j' is an index over buffer species. Equation (1) using fickian diffusion can be stated as [5, 9, 11, 14]:

$$\frac{\partial [Ca^{2+}]}{\partial t} = D_{Ca} \nabla^2 [Ca^{2+}] + \sum_j R_j \quad (2)$$

$$\frac{\partial [B_j]}{\partial t} = D_{B_j} \nabla^2 [B_j] + R_j \quad (3)$$

$$\frac{\partial [CaB_j]}{\partial t} = D_{CaB_j} \nabla^2 [CaB_j] - R_j \quad (4)$$

---

- A.Tripathi Author is with the S. V. National Institute Technology, Surat, 395007.
- N.Adlakha Author is with the S. V. National Institute Technology, Surat, 395007.

$$R_j = -k_j^+[B_j][Ca^{2+}] + k_j^-[CaB_j] \qquad (5)$$

$D_{Ca}$, $D_{B_j}$, $D_{CaB_j}$ are diffusion coefficients [1] of free calcium, free buffer, and $Ca^{2+}$ bound buffer, respectively; $k_j^+$ and $k_j^-$ are association and dissociation rate constants for buffer 'j' respectively. For stationary immobile buffers or fixed buffers $D_{B_j} = D_{CaB_j} = 0$. Given equation (1-5) can be further simplified [10, 11, 14].

$$\frac{\partial[Ca^{2+}]}{\partial t} = D_{Ca}\frac{\partial^2[Ca^{2+}]}{\partial x^2} - k_j^+[B]_\infty([Ca^{2+}]-[Ca^{2+}]_\infty) + \sigma\delta(r) \qquad (6)$$

Incorporating $j_{RYR}$, $j_{LEAK}$ and $j_{SERCA}$ in equation (6) for steady state case is given by [2, 3]:

$$\frac{d^2[Ca^{2+}]}{dx^2} - \frac{k_j^+[B]_\infty}{D_{Ca}}([Ca^{2+}]-[Ca^{2+}]_\infty) + \frac{1}{D_{Ca}}(j_{RYR} + j_{LEAK} - j_{SERCA}) = 0 \qquad (7)$$

Where

$$j_{RYR} = v_{RYR} p_o . ([Ca^{2+}]_{ER} - [Ca^{2+}])$$

$$j_{LEAK} = v_{LEAK} . ([Ca^{2+}]_{ER} - [Ca^{2+}])$$

and

$$j_{SERCA} = v_{SERCA} . \frac{[Ca^{2+}]^2}{k_{SERCA}^2 + [Ca^{2+}]^2}$$

$$\frac{d^2[Ca^{2+}]}{dx^2} - \frac{k_j^+[B]_\infty}{D_{Ca}}([Ca^{2+}]-[Ca^{2+}]_\infty) + \frac{1}{D_{Ca}} v_{RYR} p_o.([Ca^{2+}]_{ER}-[Ca^{2+}])$$

$$-\frac{1}{D_{Ca}} v_{SERCA} \frac{[Ca^{2+}]^2}{[Ca^{2+}]^2 + k_{SERCA}^2} = 0 \qquad (8)$$

The point source of calcium is assumed at the first node. Thus the appropriate boundary condition can be taken as [12, 15, 16]:

$$\lim_{x \to 0}\left(-D_{Ca}\frac{d[Ca^{2+}]}{dx}\right) = \sigma_{Ca} \qquad (9)$$

It is assumed that background concentration of $[Ca^{2+}]$ is 0.1µM and as it goes far away from the source, we have following boundary condition [9, 11, 14, 15].

$$\lim_{x \to 5}[Ca^{2+}] = 0.1\mu M \qquad (10)$$

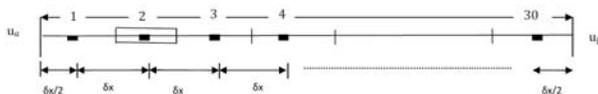

Fig 1 Discretization the domain [20]

Here, $[Ca^{2+}]_\infty$ is a background $[Ca^{2+}]$ concentration, $[B]_\infty$ is the total buffer concentration, $\sigma_{Ca}$ represents the flux due to $[Ca^{2+}]$. As stated in equation (10), $[Ca^{2+}]$ tends to the background concentration of 0.1µM as x→5.

**Case I:**
When $k_{SERCA} >> [Ca^{2+}]$ then we have

$$\frac{[Ca^{2+}]^2}{[Ca^{2+}]^2 + k_{SERCA}^2} \leq \frac{[Ca^{2+}]^2}{k_{SERCA}^2} \leq \frac{[Ca^{2+}]}{k_{SERCA}}$$
$$(a)$$

In view of above the equation (8) is taken as:

$$\frac{d^2[Ca^{2+}]}{dx^2} - \frac{k_j^+[B]_\infty}{D_{Ca}}([Ca^{2+}]-[Ca^{2+}]_\infty) + \frac{1}{D_{Ca}} v_{RYR} p_o.([Ca^{2+}]_{ER}-[Ca^{2+}])$$

$$+\frac{1}{D_{Ca}} v_{LEAK}.([Ca^{2+}]_{ER}-[Ca^{2+}]) - \frac{1}{D_{Ca}} v_{SERCA} \frac{[Ca^{2+}]}{k_{SERCA}} = 0 \qquad (11)$$

Simplifying equation (11) we get

$$\frac{d^2[Ca^{2+}]}{dx^2} - \mu[Ca^{2+}] + \lambda = 0 \qquad (12)$$

Where

$$\mu = \frac{1}{D_{Ca}}\left(k_j^+[B]_\infty + v_{RYR} p_o + v_{LEAK} + \frac{v_{SERCA}}{k_{SERCA}}\right)$$

and

$$\lambda = \frac{1}{D_{Ca}}\left(k_j^+[B]_\infty[Ca^{2+}]_\infty + v_{RYR} p_o[Ca^{2+}]_{ER} + v_{LEAK}[Ca^{2+}]_{ER}\right)$$

Integration of the equation (12) over a control volume gives [8, 20]:

$$\int_{\Delta V}\frac{d^2[Ca^{2+}]}{dx^2}dV - \int_{\Delta V}(\mu[Ca^{2+}]-\lambda)dV = 0 \qquad (13)$$

which leads to following form:

$$\left[\left(A\frac{d[Ca^{2+}]}{dx}\right)_e - \left(A\frac{d[Ca^{2+}]}{dx}\right)_w\right] - \left[(\mu[Ca^{2+}]-\lambda)A\delta x\right] = 0 \qquad (14)$$

A formula for nodal values to the east and west is developed for nodal values 2, 3, 4............29 by introducing usual linear approximation for concentration gradient, subsequent by dividing (14) by cross sectional area A and replacing the calcium concentration $[Ca^{2+}]$ is replaced by u for convenience [12, 13]. We have

$$\left[\left(\frac{u_E - u_P}{\delta x}\right) - \left(\frac{u_P - u_W}{\delta x}\right)\right] - \left[(\mu u_P - \lambda)\delta x\right] = 0 \qquad (15)$$

This (15) can be re-arranged as:

$$\left(\frac{1}{\delta x} + \frac{1}{\delta x}\right)u_P = \left(\frac{1}{\delta x}\right)u_W + \left(\frac{1}{\delta x}\right)u_E + \lambda\delta x - \mu u_P \delta x = 0 \qquad (16)$$

The general form for the interior nodal point 2, 3, 4........29 is given by:

$$a_P u_P = a_W u_W + a_E u_E + S_u \qquad (17)$$

Where $a_W = \frac{1}{\delta x}$, $a_E = \frac{1}{\delta x}$, $a_P = a_W + a_E - S_P$, $S_P = -\mu\delta x$ and

$$S_u = \lambda \delta x \qquad (18)$$

The boundary conditions are applied at nodal points 1 and 30. At node 1 west control volume boundary is kept at specified concentration and thus we get [20].

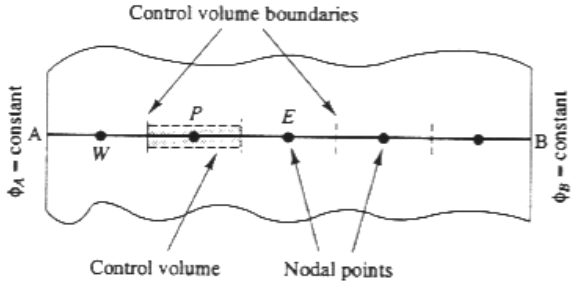

Fig 2 Control volume [20]

$$a_W = 0, \; a_E = \frac{1}{\delta x}, \; S_P = -\left(\frac{3}{\delta x} + \mu \delta x\right) \text{ and } S_u = \lambda \delta x + \frac{2u_\alpha}{\delta x} \quad (19)$$

Similarly at node 30 east control volume boundary is at specified concentration [20] and we get:

$$a_E = 0, \; a_W = \frac{1}{\delta x}, \; S_P = -\left(\frac{3}{\delta x} + \mu \delta x\right) \text{ and } S_u = \lambda \delta x + \frac{2u_\beta}{\delta x} \quad (20)$$

In equation (16), substituting all values for equation (18-20) we get the system of linear algebraic equations as given below:

$$[A]_{30\times 30} [\bar{u}]_{30\times 1} = [B]_{30\times 1} \qquad (21)$$

The Gaussian Elimination Method has been used to obtain the solution for entire problem. Here $\bar{u} = u_1, u_2, \ldots, u_{30}$ represents the calcium concentration, $A$ are system matrices and $B$ is the system vector. A computer program has been developed in MATLAB 7.11 for the problem and simulated on Core i3 processor with 2.13 GHz processing speed. 64-bit machine with 320 GB memory.

**Case II:**

When $k_{SERCA} << [Ca^{2+}]$ then we assume

$$k_{SERCA} = \alpha [Ca^{2+}] \qquad \text{for } 0 \leq \alpha \leq 1$$

$$\frac{[Ca^{2+}]^2}{[Ca^{2+}]^2 + k_{SERCA}^2} = \frac{1}{\alpha^2 + 1} \qquad (b) \qquad \text{for } 0 \leq \alpha \leq 1$$

The equation (8) is taken in the form:

$$\frac{d^2[Ca^{2+}]}{dx^2} - \frac{k_j^+[B]_\infty}{D_{Ca}}([Ca^{2+}] - [Ca^{2+}]_\infty) + \frac{1}{D_{Ca}} v_{RYR} p_o \cdot ([Ca^{2+}]_{ER} - [Ca^{2+}])$$
$$+ \frac{1}{D_{Ca}} v_{LEAK} \cdot ([Ca^{2+}]_{ER} - [Ca^{2+}]) - \frac{1}{D_{Ca}} v_{SERCA} \frac{1}{\alpha^2 + 1} = 0$$
$$(22)$$

Simplifying equation (22), we get

$$\frac{d^2[Ca^{2+}]}{dx^2} - \upsilon [Ca^{2+}] + \eta = 0 \qquad (23)$$

Where

$$\upsilon = \frac{1}{D_{Ca}}\left(k_j^+[B]_\infty + v_{RYR} p_o + v_{LEAK}\right)$$

and

$$\eta = \frac{1}{D_{Ca}}\left(k_j^+[B]_\infty [Ca^{2+}]_\infty + v_{RYR} p_o [Ca^{2+}]_{ER} + v_{LEAK}[Ca^{2+}]_{ER} - v_{SERCA}\frac{1}{\alpha^2+1}\right)$$

Similarly, integration of the equation (23) over a control volume given by [8, 20]:

$$\int_{\Delta V} \frac{d^2[Ca^{2+}]}{dx^2} dV - \int_{\Delta V} (\upsilon[Ca^{2+}] - \eta) dV = 0 \qquad (24)$$

Which leads to following form:

$$\left[\left(A\frac{d[Ca^{2+}]}{dx}\right)_e - \left(A\frac{d[Ca^{2+}]}{dx}\right)_w\right] - \left[(\upsilon[Ca^{2+}] - \eta) A \delta x\right] = 0 \quad (25)$$

A formula for nodal values to the east and west is developed for nodal values 2, 3, 4,.........29 by introducing usual linear approximation for concentration gradient. Subsequently dividing by cross sectional area A and replacing the calcium concentration $[Ca^{2+}]$ is replaced by u for convenience [12, 13].

$$\left[\left(\frac{u_E - u_P}{\delta x}\right) - \left(\frac{u_P - u_W}{\delta x}\right)\right] - \left[(\upsilon u_P - \eta)\delta x\right] = 0 \qquad (26)$$

This can be re-arranged by:

$$\left(\frac{1}{\delta x} + \frac{1}{\delta x}\right)u_P = \left(\frac{1}{\delta x}\right)u_W + \left(\frac{1}{\delta x}\right)u_E + \eta \delta x - \upsilon u_P \delta x = 0 \quad (27)$$

The general form for the interior nodal point 2, 3, 4........29 is given by:

$$a_P u_P = a_W u_W + a_E u_E + S_u \qquad (28)$$

Where

$$a_W = \frac{1}{\delta x}, \; a_E = \frac{1}{\delta x}, \; a_P = a_W + a_E - S_P, \; S_P = -\upsilon \delta x \text{ and } S_u = \eta \delta x$$
$$(29)$$

Now the boundary conditions are applied at nodal points 1 and 30. At node 1 west control volume boundary is kept at specified concentration [20].

$$a_W = 0, \; a_E = \frac{1}{\delta x}, \; S_P = -\left(\frac{3}{\delta x} + \upsilon \delta x\right) \text{ and } S_u = \eta \delta x + \frac{2u_\alpha}{\delta x}$$
$$(30)$$

Similarly at node 30 east control volume boundary is at specified concentration [20].

$$a_E = 0, \; a_W = \frac{1}{\delta x}, \; S_P = -\left(\frac{3}{\delta x} + \upsilon \delta x\right) \text{ and } S_u = \eta \delta x + \frac{2u_\beta}{\delta x} \quad (31)$$

In equation (27), substituting all values from equations (29-31) and get the system of linear algebraic equations as

given below:

$$[A]_{30\times30}[\bar{u}]_{30\times1} = [B]_{30\times1} \quad (32)$$

The Gaussian Elimination Method has been used to obtain the solution. Here $\bar{u} = u_1, u_2, \ldots, u_{30}$ represents the calcium concentration, $A$ are system matrices and $B$ is the system vector. It was stated identically by previous section.

## 3 RESULTS AND DISCUSSION

The numerical values of physical and physiological parameters used for computation of numerical results are given in Table I:

**Table I: Values of physiological parameters used for numerical results [3]**

| Symbol | Parameter | Value |
|---|---|---|
| $D_{Ca}$ | Diffusion Coefficient | 250 $\mu m^2$/sec |
| $k^+$ (EGTA) | Buffer association rate (Exogenous buffer) | 1.5 $\mu M^-$/sec |
| $k^+$ (BAPTA) | Buffer association rate (Exogenous buffer) | 600 $\mu M^-$/sec |
| $v_{RYR}$ | RYR Receptor rate | 0.5 $\mu M$/sec |
| $v_{SERCA}$ | Serca pump rate | 400 $\mu M$/sec |
| $v_{LEAK}$ | Leak rate | 0.2 $\mu M$/sec |
| $k_{SERCA}$ | Dissociation rate of SERCA | 0.2 $\mu M$ |
| $[B_m]_\infty$ | Buffer Concentration | 50 $\mu m$ |
| $[Ca^{2+}]_\infty$ | Background $Ca^{2+}$ Concentration | 0.1 $\mu m$ |
| $\sigma$ | Source amplitude | 1 pA |

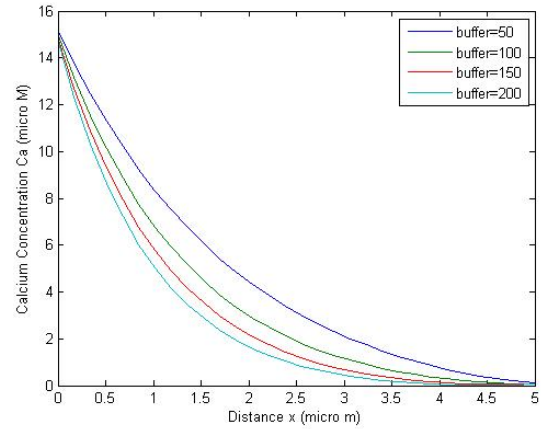

Fig 3(b) Spatial Calcium Distribution for different values of EGTA buffer and Case II

Fig. 3a and Fig. 3b shows the spatial calcium distribution in neuron cell in presence of exogenous buffer for EGTA for case I and case II respectively. In fig 3a, the $Ca^{2+}$ concentration is higher near the source along the x-direction. The $Ca^{2+}$ concentration falls down sharply between x=0 µm to x=0.4 µm, and then falls down gradually between x=0.4 µm to x=1.4 µm and finally converges its minimum value of $Ca^{2+}$ profile 0.1 µM after x=1.5 µm onwards. In all these cases the fall in $Ca^{2+}$ concentration is almost same for all values of free buffer concentration. In fig 3b, the $Ca^{2+}$ concentration is maximum near the source and it is 15 µM. The $Ca^{2+}$ concentration falls down smoothly between x=0 µm to x=0.2 µm and then fall down gradually between x=0.2 µm to x=3.5 µm then approaches its constant value 0.1 µM. In all values of buffer concentration the graph become more curvilinear shape for higher values of buffer concentration.

The difference in both cases is that in first case, when $k_{SERCA} >> [Ca^{2+}]$ the free $Ca^{2+}$ ion is minimum while in second case when $k_{SERCA} << [Ca^{2+}]$ the $Ca^{2+}$ ion in larger near the source. The comparison of fig 3(a) and fig 3(b) the removal of $Ca^{2+}$ concentration is faster in case I as compared to case II by serca pump.

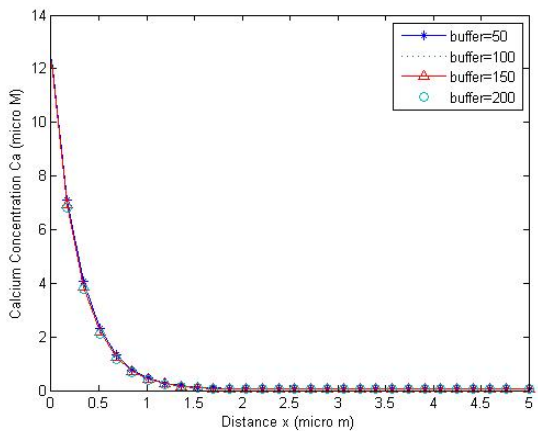

Fig 3(a) Spatial Calcium Distribution for different values of EGTA buffer and Case I

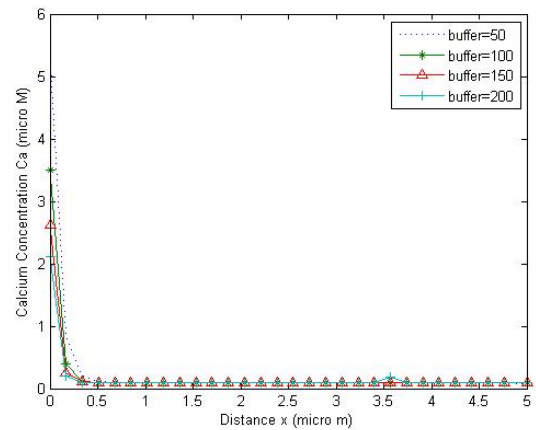

Fig 4(a) Spatial Calcium Distribution in neuron cell for different values of BAPTA buffer and Case I

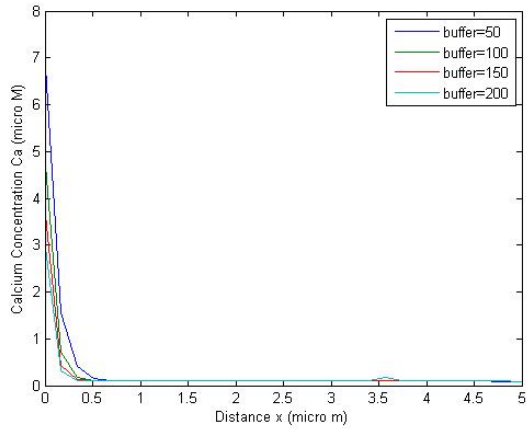

Fig 4(b) Spatial Calcium Distribution in neuron cell for different values of BAPTA buffer and Case II

Fig. 4a and fig. 4b shows the spatial variation of calcium distribution in neuro in presence of exogenous buffer BAPTA for case I and case II respectively. In fig 4a, the $Ca^{2+}$ concentration got maximum near the source along the x-direction. The $Ca^{2+}$ concentration falls down sharply between x=0 µm to x=0.3 µm, and then falls down gradually between x=0.3 µm to x=0.5 µm and finally converges its minimum value of $Ca^{2+}$ profile 0.1 µM. In all these cases the falls in $Ca^{2+}$ concentration is higher for lower value of free buffer concentration 50 µM and lower for higher value of buffer concentration 200µM. In fig 4b, the $Ca^{2+}$ concentration is higher near the source i.e. 5. 2 µM and then remain same as case I.

The $Ca^{2+}$ concentration got the maximum value for smaller value of buffer concentration its 50 µM and it gets minimum for higher values of buffer concentration i.e. 200 µM. The $Ca^{2+}$ concentration near the source for EGTA is very significantly high than that for BAPTA. The difference in results for EGTA and BAPTA is due to the fact that EGTA is slow buffer and BAPTA is fast chelator. Further the effect of SERCA pump is more significantly visible in case of EGTA buffer as compare to the cases involving BAPTA buffer.

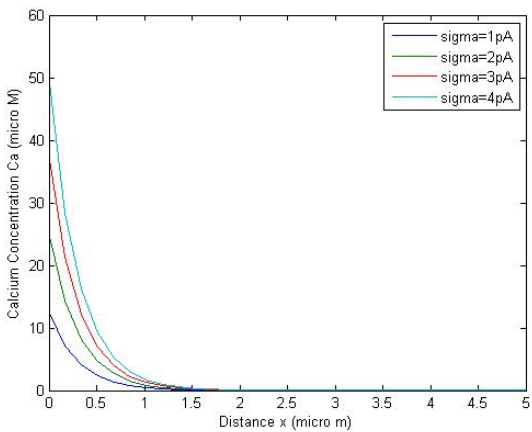

Fig 5(a) Spatial Calcium Distribution in neuron cell for EGTA buffer, with different values of sigma and Case I

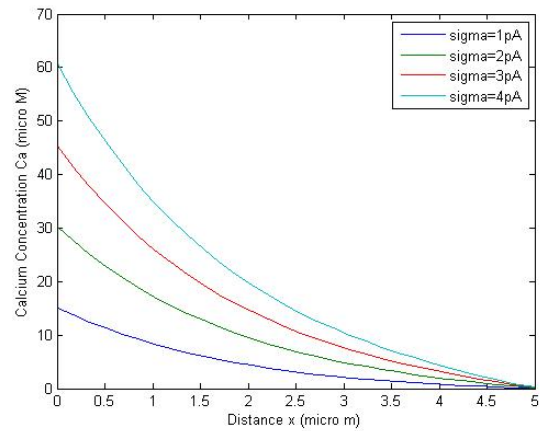

Fig 5(b) Spatial Calcium Distribution in neuron cell for EGTA buffer, with different values of sigma and Case II

Fig. 5a and fig. 5b shows the spatial variation of $Ca^{2+}$ distribution profiles for EGTA buffer with different values of flux and for case I and case II respectively. In fig 5a, the fall of $Ca^{2+}$ concentration is sharp for all values of flux and the concentration of $Ca^{2+}$ profile becomes constant after distance x=1 µm. The calcium concentration is high near the source for higher values of flux sigma. In fig 5b, the fall of $Ca^{2+}$ concentration is smooth for all values of flux sigma and the concentration of $Ca^{2+}$ profile becomes almost constant after distance x=4 µm. Also calcium concentration is high near the source for higher values of flux sigma.

Comparing figures 5a and 5b, we observe that the fall in $Ca^{2+}$ concentration is more in fig 5a than in fig 5b. This shows that there is significant difference in curves due to difference in two conditions of the serca pump. Further the gaps among the curves in both the figures 5a and 5b indicate that flux has significant affect on the calcium distribution in the neuron cell.

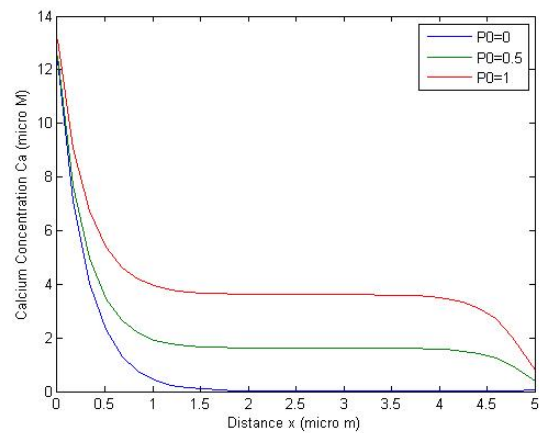

Fig 6(a) Spatial Calcium Distribution in neuron cell with EGTA buffer, different states of receptor and case I

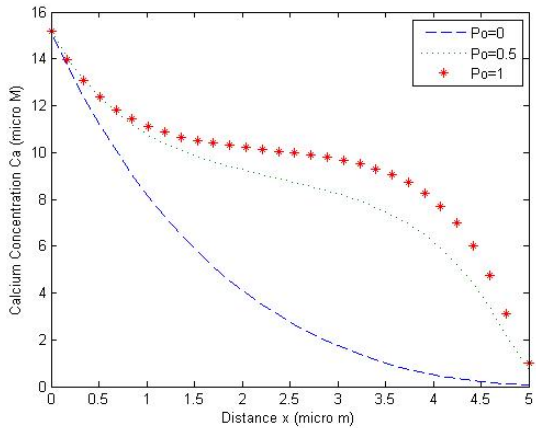

Fig 6(b) Spatial Calcium Distribution in neuron cell with EGTA buffer, different states of receptor and case II

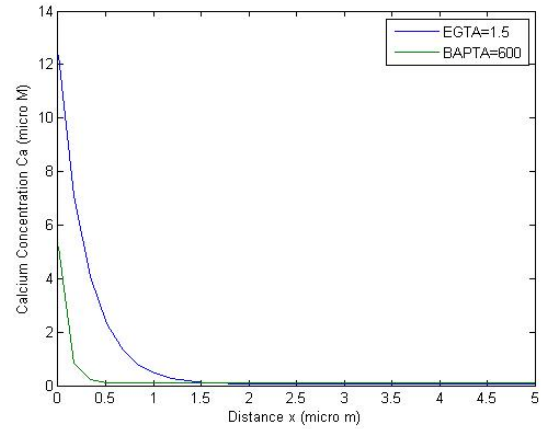

Fig 7(a) Spatial Calcium Distribution in neuron cell with different chelator EGTA and BAPTA and case I

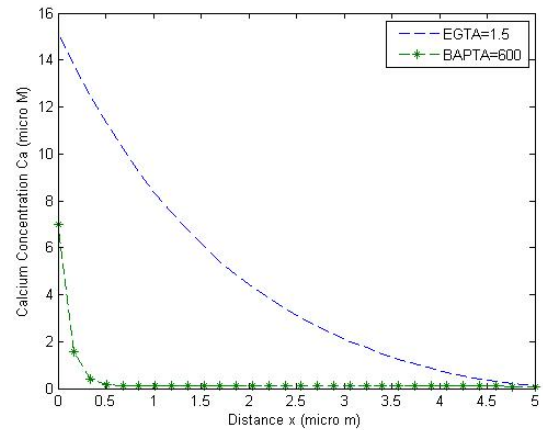

Fig 7(b) Spatial Calcium Distribution in neuron cell with different chelator EGTA and BAPTA and case II

Fig. 6a and fig. 6b are for spatial $Ca^{2+}$ distribution in neuron cell in presence of EGTA buffer with different states of receptor for case I and case II respectively. The probability $P_0=0$ represents the state when receptor channel is closed where as $P_0=1$ represents the represents the state when receptor channel is completely open. Also $P_0=0.5$ represents the state when channel is 50% open.

In fig 6a we see that fall in calcium concentration is sharp between x=0 μm and x=0.5 μm and then it becomes almost constant after x=1.5 μm. The fall in calcium concentration is more when the ryanodine channel is closed state as compared to the states where receptor is partially open and completely open. In fig 6b we observe that the fall in calcium concentration is more gradual and smooth when receptor channel is partially open and completely open as compared to the state when receptor channel is closed.

The reason is that when channel is closed the free calcium released from the cytosol and decreases rapidly. If the channel is partially open then it comes slowly for the long duration then it approaches at constant value. If the channel is completely open then it comes more slowly for long duration and becomes constant.

Thus it is evident from above that receptor has significant effect on calcium profiles in the neuron cells and this effect is more significant for case II as compared to case I.

Fig. 7a and Fig. 7b shows spatial variation of calcium concentration with space and association rate case I and case II respectively. Fig. 7a the fall of calcium concentration is smoothly decaying for EGTA as compare to the fall of $Ca^{2+}$ ion concentration sharply decreasing for BAPTA. Fig 7b as we move away from the source fall in calcium concentration with BAPTA buffer is sharp [15] whereas the fall in calcium concentration with EGTA buffer is evidently slow. The calcium concentration achieves the constant value with BAPTA while it doesn't achieve steady state with EGTA. The calcium ion concentration for EGTA is lower compared to BAPTA. This is because EGTA has a lower association rate i.e. there is lesser binding of calcium ions and higher concentration of free calcium ions in the cytosol. Therefore, there is a larger extrusion of calcium ions out of the cytosol and consequently a lesser calcium ion concentration.

## 4 CONCLUSION

Finite volume method is quite flexible, powerful and versatile in the present study, as it was possible to incorporate more realistic assumption regarding the biophysical parameters in the problem. Further the results obtained there the light on effect of serca pump, receptors, flux and buffers on spatial calcium concentration in neuron cell, which was found to be significant. Such finite volume models can be further developed for more realistic studies in higher dimension in near future to obtain information which can be of great use to biomedical scientists for developing protocols for diagnosis and treatment of neuronal diseases.

**Amrita Tripathi** is a research scholar at department of Applied Mathematics and Humanities, S. V. National Institute of Technology Surat (Gujarat). Email: t.amrita@ashd.svnit.ac.in,
 tripathi.amrita28@gmail.com

**Neeru Adlakha** is a Associate Professor at department of Applied Mathematics and Humanities, S. V. National Institute of Technology Surat (Gujarat). Email: nad@ashd.svnit.ac.in